\documentclass[a4paper]{jpconf}
\usepackage{graphicx}
\usepackage{subfig}
\usepackage{caption}
\usepackage{amsmath}
\usepackage{mathtools}
\usepackage{nicefrac}
\usepackage{cleveref}

\newcommand{\ba}{\begin{eqnarray}}
\newcommand{\ea}{\end{eqnarray}}

\begin{document}
\captionsetup[figure]{labelfont={bf},labelsep=period}
\captionsetup[table]{labelfont={bf},labelsep=period}
\title{Splitting of single-particle levels in clusters potentials}

\author{A.H. Santana-Vald\'es* and R. Bijker}
\address{Instituto de Ciencias Nucleares, 
Universidad Nacional Aut\'onoma de M\'exico, 
A.P. 70-543, 04510 M\'exico, D.F., M\'exico}
\ead{*adrian.santana@correo.nucleares.unam.mx}

\begin{abstract}
In analogy with the Nilsson model, we calculate the splitting of spherical single-particle levels in a deformed field, but for cluster potentials. We study applications to alpha-cluster nuclei with two, three and four alpha particles, in which the deformation corresponds to the relative distance between the alpha particles. The splitting of the single-particle levels is studied for the cases of a dumbbell, equilateral triangle and a regular tetrahedron. The observed patterns may be used to gain insight into how the single-particle levels evolve with deformation.
\end{abstract}

\section{Introduction}

\begin{figure}[h]
\centering{}
{\centering{}\includegraphics[scale=0.55]{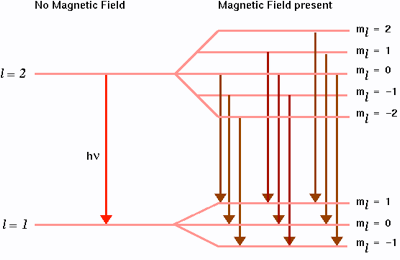} }\hspace{2pc}  \begin{minipage}[b]{16pc}\caption{Zeeman effect: splitting of levels due to an external magnetic field.
\label{fig:Zeeman effect}}
\end{minipage}

\end{figure}

\begin{figure}[h]
\centering{}
{\centering{}\includegraphics[scale=0.55]{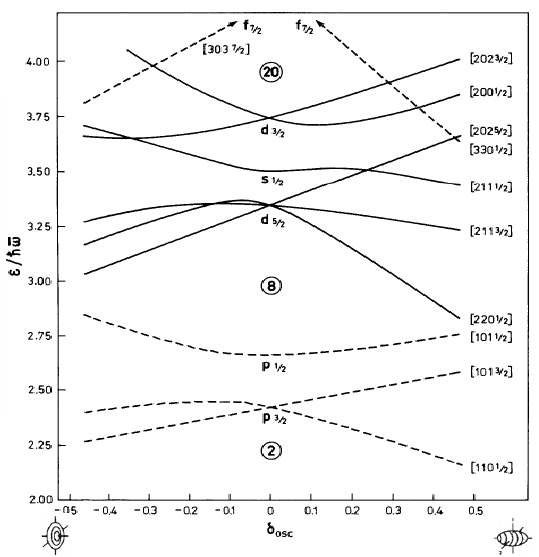}}\hspace{2pc}  \begin{minipage}[b]{15pc}\caption{Splitting of single-particle levels as a function of the quadrupole deformation in the deformed shell model. Levels with even and odd parity are drawn with solid and dashed lines respectively (taken from \cite{bohr1998nuclear}).
\label{fig:Nilsson Model}}
\end{minipage}

\end{figure}

Energy level splitting is a very well-known phenomenon in physics, which occurs whenever a set of degenerate levels is split in an external field. One of the best known examples is the Zeeman effect in which rotational states are split in an external magnetic field according to their magnetic substates (\Cref{fig:Zeeman effect}) \cite{bohr1998nuclear}.
\\
Another example is the Nilsson model of deformed shell model where the single-particle levels are split in a deformed quadrupole field \cite{casten2000nuclear}. In this case the energy levels are not only split but also strongly mixed by the quadrupole field. The amount of mixing depends on the strength of the deformed field (\Cref{fig:Nilsson Model}). Level crossing and repulsion occur due to this deformation.
In the case of the Zeeman effect, the levels are split but not mixed, and can therefore be labeled by the same quantum numbers as before the splitting, i.e the angular momentum $l$ and its projection m$_{l}$. In the case of the Nilsson model, the levels are strongly mixed and are labeled by the quantum numbers in the intrinsic or body-fixed frame $\left[\mathrm{N}n_{z}\varLambda\mathrm{\Omega}\right]^\pi$ where $\Omega$ denotes the projection of the single-particle angular momentum on the symmetry axis (z-axis), $\pi$ is the parity, $N$ the principal quantum number, $n_{z}$ the number of nodes in the wave function in the z-direction and $\varLambda$ the projection of the orbital angular momentum on the symmetry axis. 
\\
In our work, we use a deformed field, that of cluster potentials, where the clusters are alpha particles.
The deformation is taken to be the relative distance to the center of mass of the whole $k\alpha+x$ nucleon structure, where is $x$
a nucleon. The splitting of the single-particle levels is studied
as a function of the relative distance for the cases of a dumbbell,
equilateral triangle and a regular tetrahedron, where a symmetry factor
is crucial for the level splitting and configuration mixing. Relabeling
is also necessary in this case specially for the cases of an equilateral
triangle and regular tetrahedron. The main goal for this study is
to obtain patterns that gain insight into how the single-particle levels evolve with deformation. That is done by studying the mixing between states, how these are related with the geometrical configuration and what kind of symmetry is present. 

\section{Cluster Shell Model}
We start by reviewing the Cluster Shell Model \cite{DELLAROCCA2017158}. In this model nuclei with $Z=N=2k$ are treated as a cluster of $k$ $\alpha$-particles whose matter and charge density are given by a gaussian form
\begin{align}
\rho\left(\vec{r}\right)=\left(\frac{\alpha}{\pi}\right)^{\frac{3}{2}}e^{-\alpha\left(r^{2}+\beta^{2}\right)}4\pi\sum_{\lambda\mu}i_{\lambda}\left(2\alpha\beta r\right)Y_{\lambda\mu}\left(\theta,\phi\right)\sum_{i=1}^{k}Y_{\lambda\mu}^{*}\left(\theta_{i},\phi_{i}\right).\label{eq:distributionnucleus}
\end{align}
Here $\vec{r}_{i}=\left(\beta,\theta_{i},\phi_{i}\right)$ where $\beta$ denotes the relative distance of the $\alpha$ particles to the center of mass, and $\theta_{i}$ and $\phi_{i}$ are the angles. An important factor is the cluster factor, $\sum_{i=1}^{k}Y_{\lambda\mu}^{*}\left(\theta_{i},\phi_{i}\right)$, which contains the information about the geometrical configuration of the alpha particles.
\\
The potential is obtained by convoluting the density, \cref{eq:distributionnucleus}, with a Volkov potential \cite{VOLKOV196533} to obtain
\begin{equation}
V\left(\vec{r}\right)=-V_{0}\sum_{\lambda\mu}4\pi e^{-\alpha\left(r^{2}+\beta^{2}\right)} i_{\lambda}\left(2\alpha\beta r\right)Y_{\lambda\mu}\left(\theta,\phi\right)\sum_{i=1}^{k}Y_{\lambda\mu}^{*}\left(\theta_{i},\phi_{i}\right)\label{eq:potencialcentral}.
\end{equation} 
\\
\begin{center}
\begin{figure}[h]
\centering{}\includegraphics[scale=0.5]{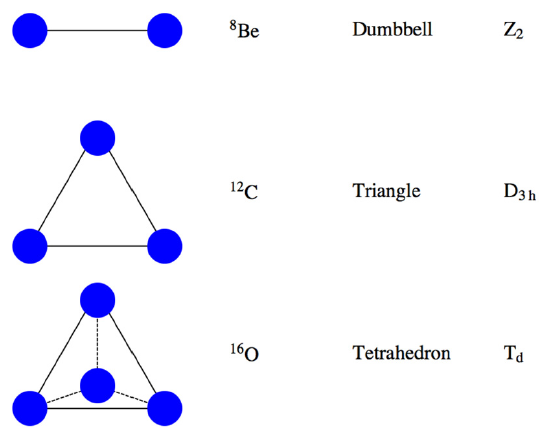}\hspace{2pc}\begin{minipage}[b]{15pc}\caption{\label{fig:Geometrical-configuration}Geometrical configuration for $k=2,3,4$ clusters (taken from \cite{DELLAROCCA2017158}).}
\end{minipage}

\end{figure}
\par\end{center}
The spin-orbit interaction is taken as
\begin{equation}
V_{so}\left(\vec{r}\right)=V_{0,so}\frac{1}{2}\left[\frac{1}{r}\frac{\partial V\left(\vec{r}\right)}{\partial r}\left(\vec{s}\cdot\vec{l}\right)+\left(\vec{s}\cdot\vec{l}\right)\frac{1}{r}\frac{\partial V\left(\vec{r}\right)}{\partial r}\right].
\end{equation}
Finally, the coulomb potential is found by convoluting the carge density $\frac{Ze^2}{k}\rho\left(\vec{r}\right)$ with Green's function giving 
\begin{eqnarray}
V_{C}\left(\vec{r}\right) & = & \frac{Ze^{2}}{k}\left(\frac{\alpha}{\pi}\right)^{\frac{3}{2}}\sum_{\lambda\mu}\frac{4\pi}{2\lambda+1}Y_{\lambda\mu}\left(\theta,\phi\right)\sum_{i=1}^{k}Y_{\lambda\mu}^{*}\left(\theta_{i},\phi_{i}\right)\times \\
 &  & \times\left[\frac{1}{r^{\lambda+1}}\intop_{0}^{r}f_{\lambda}\left(r'\right)r'^{\lambda+2}dr'+r^{\lambda}\intop_{r}^{\infty}f_{\lambda}\left(r'\right)\frac{1}{r'^{\lambda-1}}dr'\right].
\end{eqnarray}
We then obtain the single-particle energy levels and intrinsic states
as a function of $\beta$ for each configuration $\left(k=2,3,4\right)$
of \cref{fig:Geometrical-configuration} by solving the single-particle
Schr$\mathrm{\ddot{o}}$dinger equation
\begin{equation}
H=\frac{\vec{p}^{2}}{2m}+V\left(\vec{r}\right)+V_{so}\left(\vec{r}\right)+V_{C}\left(\vec{r}\right)\label{eq:hamiltonian free particle}.
\end{equation}
In the case of neutrons $V_{C}\left(\vec{r}\right)=0$.

\section{Results}

\begin{center}
\begin{figure}[h]
\noindent \begin{centering}
\subfloat {\includegraphics[scale=0.45]{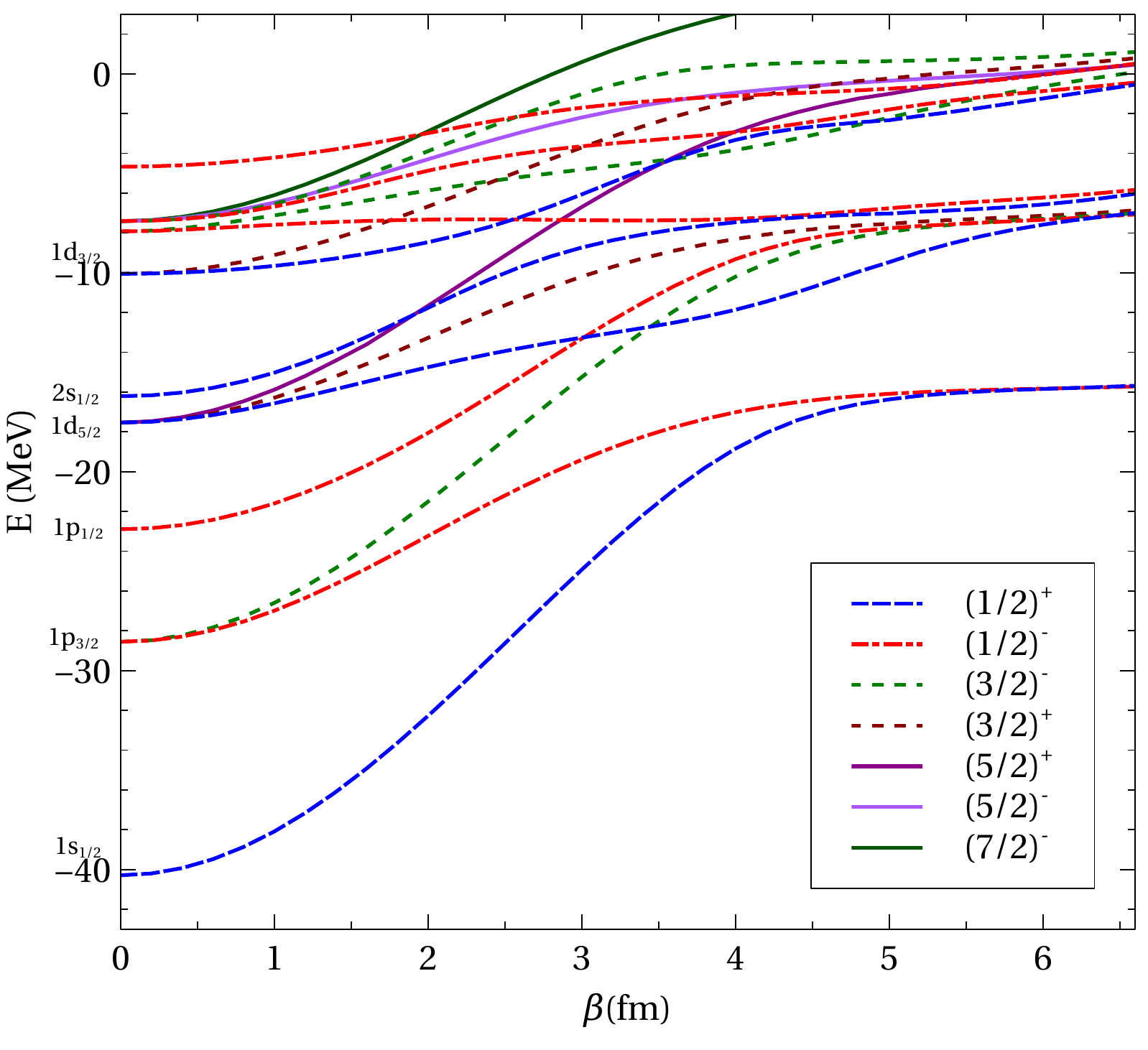}
\label{fig:Energy-level-splitting DumbbellProton}}
\qquad
\subfloat {\includegraphics[scale=0.45]{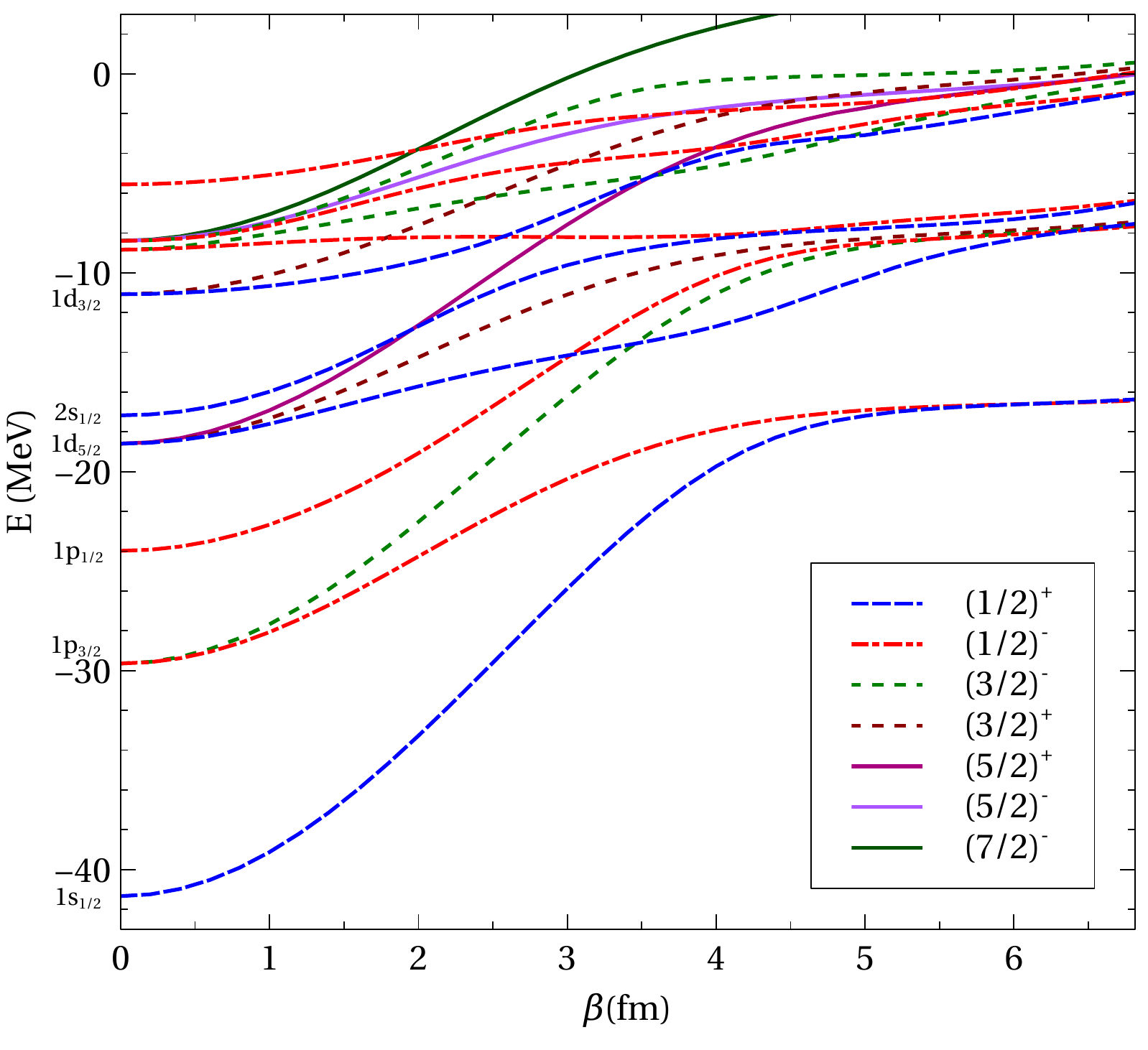}
\label{fig:Energy-level-splitting DumbbellNeutron}}
\caption{\label{Correlation-Diagrams_pvsn}Correlation diagrams for the case of dumbbell, obtained from \cref{eq:hamiltonian free particle} with $V_{0}=32$ MeV, $\alpha=0.0511\ \mathrm{fm}^{-2}$ and $V_{0,so}=27.5$ MeV fm$^2$ \cite{DELLAROCCA2017158} for a proton (left) and a neutron (right).}
\par\end{centering}
%
%
%
\end{figure}
\end{center}

\begin{figure}[!h]
\subfloat {\includegraphics[scale=0.50]{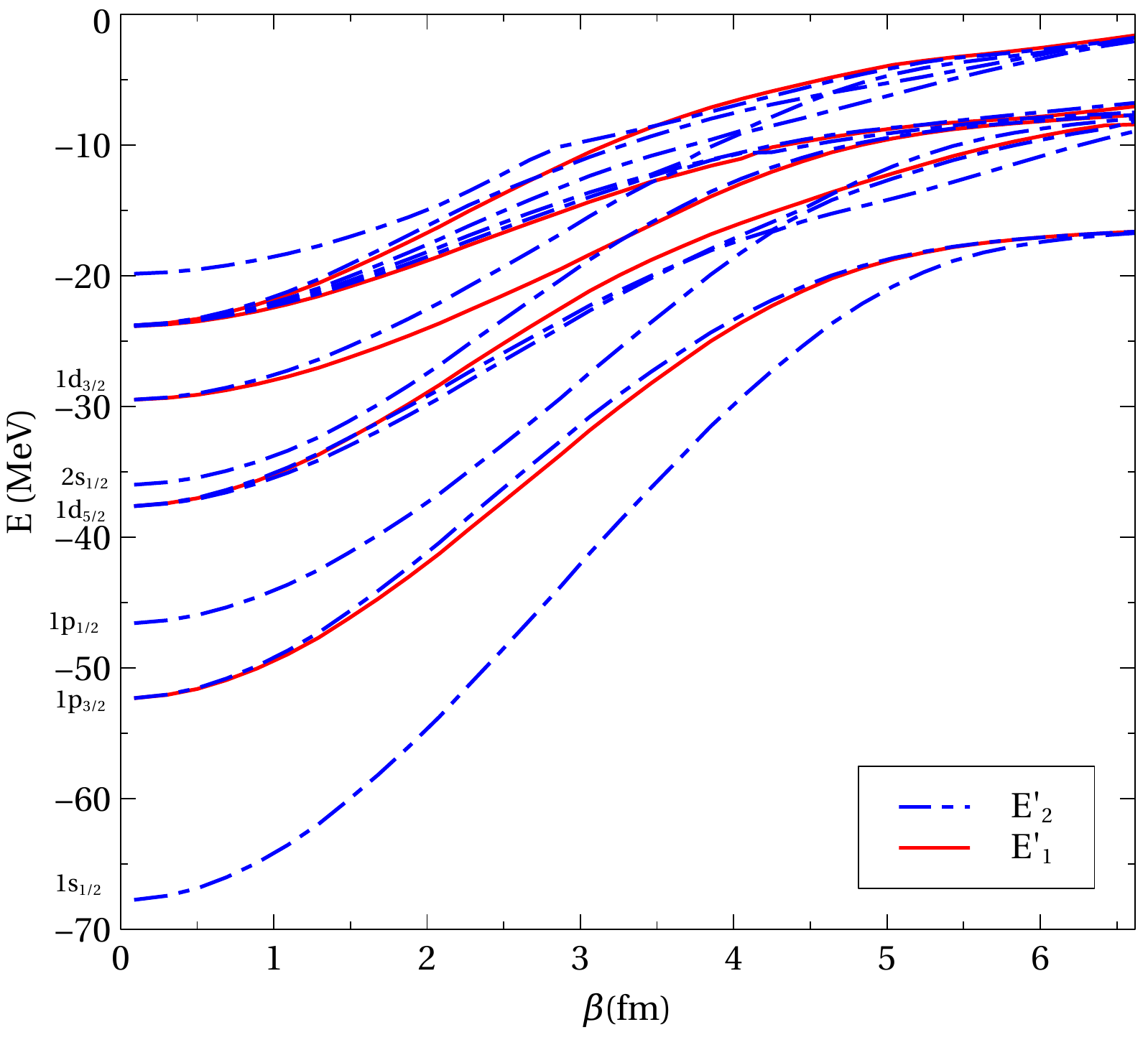}
\label{Correlationtriangle}}
\quad
\subfloat{\includegraphics[scale=0.50]{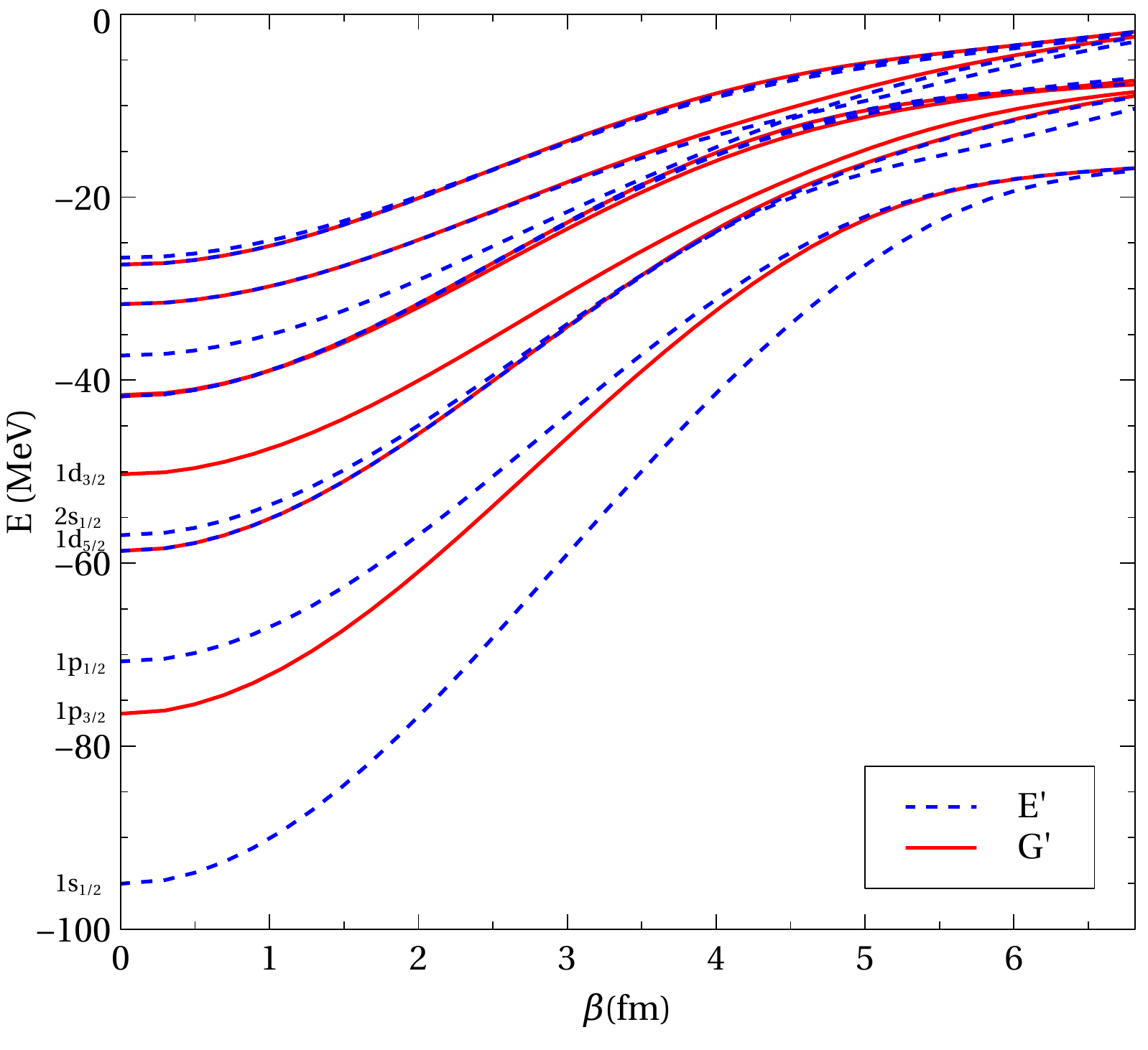}
\label{Correlationtetrahedron}}
\\
\subfloat{\includegraphics[scale=0.50]{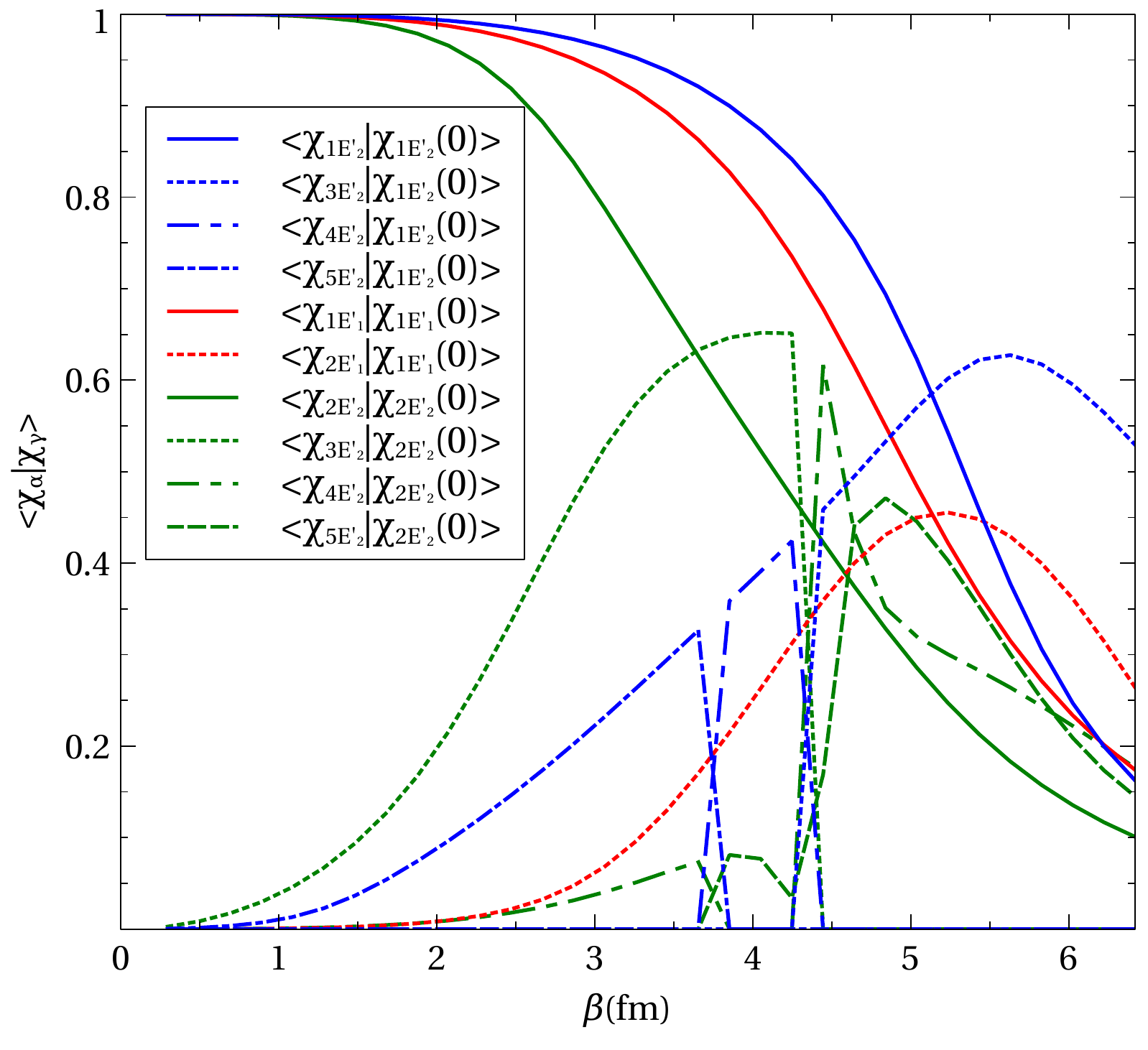}
\label{overlaptriangle}}
\quad
\subfloat{\includegraphics[scale=0.50]{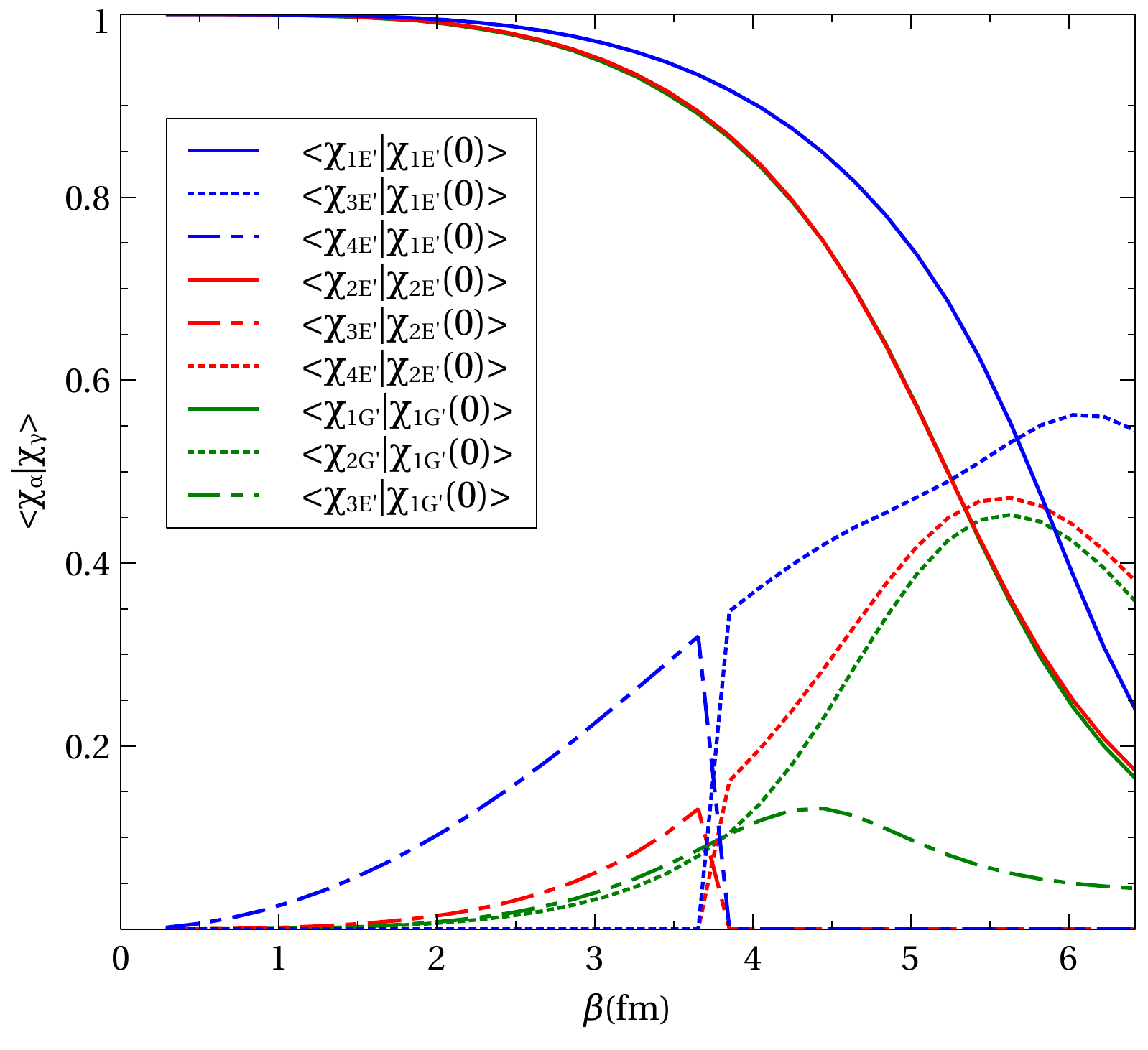}
\label{overlaptetrahedron}}
\caption{\label{CorrTrianPirOvlp}Correlation diagrams (top) and overlaps between states of the same symmetry (bottom) for the case of a triangle (left) and a tetrahedron (right), and a single-particle neutron, obtained from \cref{eq:hamiltonian free particle} with $V_{0}=32$ MeV, $\alpha=0.0511\ \mathrm{fm}^{-2}$ and $V_{0,so}=17$ MeV $\mathrm{fm^{2}}$ for the triangle, and $V_{0,so}=12.1$ MeV $\mathrm{fm^{2}}$ for the tetrahedron \cite{DELLAROCCA2017158}.}
\end{figure}

The Hamiltonian of \cref{eq:hamiltonian free particle} is diagonalized in the harmonic oscillator basis. The correlation diagrams are shown as a function of $\beta$ in \Cref{Correlation-Diagrams_pvsn,CorrTrianPirOvlp}. The states for the spherical case correspond to $\beta=0$. 
For the two-body cluster configuration the levels with projection $\pm \Omega$ are degenerate. As a consequence, a single-particle level $\left(l,\nicefrac{1}{2}\right)j$ is split into a series of doublets with $\Omega=\nicefrac{1}{2},\ \nicefrac{3}{2},\ \ldots,\ j $ and parity $P=\left(-\right)^l$.
For the triangular configuration neither the angular momentum nor parity is conserved. The spherical single-particle levels are split into a series of doublets characterized by $E'_{2}$ and $E'_{1}$, the double valued representations of $D_{3}$, the rotational subgroup of $D_{3h}.$ 
For the tetrahedral configuration, the single-particle levels are split into a series of doublets and quadruplets characterized by $E'$ and $G'$, respectively, the double valued representations of $T$, the rotational subgroup of $T_{d}$. The results are summarized in \cref{EnergyLvlsplit}.
\begin{table}[h]
\centering
\caption[]{Resolution of single-particle levels into the double-valued representations of the corresponding point group. Notation from Hamermesh \cite{hamermesh1962group}.}
\label{EnergyLvlsplit}
\begin{tabular}{cccc}
\br
\noalign{\smallskip}
$j$ & Dumbbell & Triangle & Tetrahedron \\
\noalign{\smallskip}
\hline
\noalign{\smallskip}
$\nicefrac{1}{2}$ & $(\nicefrac{1}{2})^P$  & $E'_{2}$ & $E'$ \\
$\nicefrac{3}{2}$ &$(\nicefrac{1}{2}+ \nicefrac{3}{2})^P$ & $E'_{1}+E'_{2}$  & $G'$ \\
$\nicefrac{5}{2}$ &$(\nicefrac{1}{2}+\nicefrac{3}{2}+
\nicefrac{5}{2})^P$ & $E'_{1}+2E'_{2}$ &  $E'+G'$ \\  
$\nicefrac{7}{2}$ &$(\nicefrac{1}{2}+\nicefrac{3}{2}+ \nicefrac{5}{2}+\nicefrac{7}{2})^P$ & $E'_{1}+3E'_{2}$  & $2E'+G'$          \\ 
\noalign{\smallskip}
\br
\end{tabular}
\end{table}
\\
The first result that we can see by comparing \Cref{Correlation-Diagrams_pvsn} is that the correlation diagrams for protons and neutrons behave in the same manner. This result shows that it does not matter much whether the particle is a neutron or a proton, there is just a small shift in energy. 
\\
In order to adress the issue of mixing between the levels we study the overlap between intrinsic states of different values of deformation, $\left\langle \chi_{\Omega}\left(\beta\right)\left|\chi_{\Omega'}\left(\beta'\right)\right.\right\rangle $. We can then have the overlap with a fixed value of $\beta'$, namely $\beta'=0$. In that way we can see the mixing between states as we vary $\beta$. This can be seen in \Cref{CorrTrianPirOvlp}, where from comparing the graphs of overlap of the triangle and tetrahedron, there is present more mixing in the case of triangle than the case of tetrahedron. A main reason why this occurs is because the density of states of same symmetry is higher in the configuration of triangle, as a result those levels are then closer energetically to each other and can mix more strongly in comparison to the tetrahedron.

For the case of triangle there exist strong repulsion between states, which causes crossings of wave function from one level to another. These cases can be seen in the graph of overlap of the triangle in the region 3.5 to 4.5 fm for the states belonging to the $E'_{2}$ symmetry. In states belonging to the $E'_{1}$ that doesn't happen, and the mixing occurs smoothly with no wave function crossings. 
\\
In the case of the tetrahedron, although there occurs a wave function crossing (between 3.5 to 4 fm) this doesn't happens as much as in the case of triangle. There is strong mixing but states are sufficiently separated so that wave function crossings don't occur as often as in the case of triangle.
\\
In both cases the difference between states of the same symmetry group arise from how much energetically distanced they are from other states. A simple example are the ground levels, which in comparison to higher levels present much less mixing. That can be seen by comparing the curves $\left\langle \chi_{\Omega}\left(\beta\right)\left|\chi_{\Omega}\left(\beta'=0\right)\right.\right\rangle$ where $\Omega$ can be any of the double-valued representations for triangle and tetrahedron shown in \Cref{EnergyLvlsplit}. The faster those curves decrease the more the state has mixed with others. Again, the mayor difference between the states is seen in the case of the triangle. Also the higher the state of the same group of symmetry the more it mixes.
\\
From both correlation diagrams of \Cref{CorrTrianPirOvlp} it can be seen that  the single-particle levels asymptotically tend to a certain degeneracy, which corresponds to the case where the alpha particles are too far apart and thus interact as separated alpha particles. This is something that is not obtained in the Nilsson model, where the degeneracy shown is that of an elongated cigar. As this occurs mixing between states decrease. This is shown in the curves of $\left\langle \chi_{\Omega}\left(\beta\right)\left|\chi_{\Omega'}\left(\beta'=0\right)\right.\right\rangle$. Here $\Omega$ and $\Omega'$ are states of the same symmetry, but not the same state. The decrease starts in most cases at 5 fm, hinting that the frontier between the moment where the alphas can still be seen as to be part of the same nucleus and not separate particles is around that value. In the correlation diagrams it is also shown that convergence towards the degeneracy of separate alphas occur after 5 fm.  
\section{Conclusions}
From the obtained patterns it has been illustrated how single-particle
levels evolve with deformation and which states interact most and
why. From the overlap it can also be thought which interactions could
be the predominant ones (quadrupole, octupole, etc). This and the
interpretation of the behaviour due to the symmetry will be further
analysed. All the information from this study can later be used to
extend the Algebraic Cluster Model (ACM)\cite{Bijker2000}\cite{Bijker2014} to odd-cluster nuclei like
$^{9}$Be, $^{9}$B, $^{13}$C, $^{13}$N, $^{17}$O and $^{17}$F,
by developing the Algebraic Cluster-Fermion Model (ACFM)\cite{Adrian2018}. 

\ack

This work was supported in part by grant IN109017 from DGAPA-UNAM, Mexico.


\section*{References}

\begin{thebibliography}{99} 
\bibitem{bohr1998nuclear}
Bohr A and Mottelson B R 1998 {\it Nuclear Structure} vol 2 (World Scientific)

\bibitem{casten2000nuclear}
Casten R 2000 {\it Nuclear Structure from a Simple Perspective} (Oxford University Press)

\bibitem{DELLAROCCA2017158}
Della Rocca V , Bijker R and Iachello F 2017 {\it Nucl. Phys.} A {\bf 966} 158-84

\bibitem{VOLKOV196533}
Volkov A B 1965 {\it Nucl. Phys.} {\bf 74} 33-58 

\bibitem{hamermesh1962group}
Hamermesh M 1962 {\it Group Theory and Its Application to Physical Problems} (Dover Publications)

\bibitem{Bijker2000}
Bijker R and Iachello F 2000 {\it Phys. Rev.} C {\bf 61} 067305

\nonum
Bijker R and Iachello F 2002 {\it Ann. Phys.} (N.Y.) {\bf 298} 334

\bibitem{Bijker2014}
Bijker R and Iachello F 2014 {\it Phys. Rev. Lett.} {\bf 112} 152501

\nonum
Bijker R and Iachello F 2017 {\it Nucl. Phys.} A {\bf 957} 154

\bibitem{Adrian2018}
Santana-Vald\'{e}s A H and Bijker R 2018 Work in progress
\end{thebibliography}
\end{document}